\documentclass[aps,twocolumn,showpacs]{revtex4}
\usepackage{graphicx}
\usepackage{amsfonts,mathptm}
\begin{document}
\title{A dynamical approach to the spatiotemporal aspects of the
Portevin-Le Chatelier effect: Chaos, turbulence and band
propagation}
\author{G. Ananthakrishna$^{1,2}$ and M.S. Bharathi$^{1}$}
\affiliation{$^1$ Materials Research Centre, Indian Institute of Science, Bangalore-560012, India\\
$^{2}$ Centre for Condensed Matter Theory, Indian Institute of Science,
Bangalore-560012, India}

\begin{abstract}

The analysis of experimental time series obtained from single and
poly-crystals subjected to a constant strain rate tests report an
intriguing dynamical crossover from a low dimensional chaotic
state at medium strain rates to an infinite dimensional power law
state of stress drops at high strain rates. We present results of
an extensive study of all aspects of  the PLC effect  within the
context a recent  model that reproduces this crossover. We
characterize the dynamics of this crossover by studying the
distribution of the Lyapunov exponents as a function of the strain
rate, with special attention  to system size effects.  The
distribution of the exponents changes from a small set of positive
exponents in the chaotic regime to a dense set of null exponents
in the scaling regime. As  the latter  is similar to the result in
the GOY shell model for turbulence, we compare the results of our
model with that of the GOY model. Interestingly, the null
exponents in our model themselves obey a power law. The study is
complimented by visualizing  the configuration of dislocations
through the slow manifold analysis. This shows that while a large
proportion of dislocations are in the pinned state in the chaotic
regime, most of them are pushed to the threshold of unpinning in
the scaling regime, thus providing an insight into the mechanism
of crossover. We also show that this model qualitatively
reproduces the different types of deformation bands seen in
experiments. At high strain rates where propagating bands are
seen, the model equations can be reduced to the Fisher-Kolmogorov
equation for propagative fronts. Marginal stability analysis shows
that the velocity of the propagating of the bands varies linearly
with the strain rate and inversely with the dislocation density.
These results are consistent with the known experimental results.
We also discuss the connection between the nature of band types
and the dynamics in the respective regimes. The analysis
demonstrates that this simple dynamical model captures  the
complex spatio-temporal features of the PLC effect.
\end{abstract}
\pacs{05.65.+b, 05.45.Ac, 62.20.Fe, 05.90.+m}
\maketitle

\section{INTRODUCTION}

Plastic deformation is a highly dissipative irreversible
nonequilibrium process where nonlinearities play a fundamental
role. Under normal conditions, one finds homogeneous deformation.
However,  under suitable conditions of deformation, different
types of spatial and temporal patterns are observed. At a
microscopic level, these patterns arise due to the collective
behavior of dislocations. These  can be broadly classified by the
associated time and length scales. For example, the persistent
slip bands observed in cyclic deformation is an example of nearly
permanent pattern lasting over long time scales \cite{KFA}. On the
other hand, a type of propagative bands referred to as the
L\"uders bands observed in uniaxial tension tests are
characterized by short time scale \cite{KFA}. Yet another and even
more complex spatio-temporal patterns are  observed during tension
tests of dilute metallic alloys in a certain range of strain rates
and temperatures. This phenomenon has come to be known as the
Portevin-Le Chatelier (PLC) effect \cite{PLC}. Here a uniform
deformation mode becomes unstable leading to a spatially and
temporally inhomogeneous state. The instability manifests itself
in the form of serrations on the stress-strain curves of the
sample \cite{KFA,HZ}. Each stress drop is generally associated
with the nucleation and often the propagation of a band of
localized plastic deformation. In poly-crystals, these bands and
the associated serrations are classified into three generic types.
On increasing strain rate or decreasing the temperature, one first
finds the type C band, identified with randomly nucleated static
bands with large characteristic stress drops on the stress- strain
curve. The serrations  are quite regular. Then the type B
'hopping' bands are seen. The serrations are more irregular with
amplitudes that are smaller than that for  the type C. The bands
formed are still localized and static in nature, but forming ahead
of the previous band in a spatially correlated way giving the
visual impression of a hopping propagation. Finally, one observes
the continuously propagating type A bands associated with small
stress drops. (In single crystals such a clear classification does
not exist.) These different types of PLC bands are believed to
represent distinct correlated states of dislocations in the bands.

The well accepted classical explanation of the PLC effect is via
the  dynamic strain aging (DSA) concept first introduced by
Cottrell \cite{Cottrell} and later extended by others
\cite{KFA,van,penning,Kubin85}. In the Cotrell's picture, the
dynamic strain aging refers to the interaction of mobile
dislocations with the diffusing solute atoms. At low strain rates
(or high temperatures) the average velocity of dislocations is low
and there is sufficient time for the solute atoms to diffuse to
the dislocations and pin them (called as aging). Thus, longer the
dislocations are arrested, larger will be the stress required to
unpin them.  When these dislocations are unpinned, they move at
large speeds till they are arrested again. At high strain rates
(or low temperatures), the time available for solute atoms to
diffuse to the dislocations decreases and hence the stress
required to unpin them decreases. Thus, in a range of strain rates
and temperatures where these two time scales are of the same order
of magnitude, the PLC instability manifests. The competition
between the slow rate of pinning and sudden unpinning of the
dislocations, at the macroscopic level translates into a negative
strain rate sensitivity (SRS) of the flow stress as a function of
strain rate which is the basic instability mechanism used in most
phenomenological models \cite{KFA,HZ}. Slow-fast dynamics and the
negative flow rate characteristic is common  to many stick-slip
systems such as frictional sliding \cite{Pers}, fault dynamics
\cite{Carlson} and peeling of an adhesive tape
\cite{Maug} and charge density waves \cite{Dumas}.\\

There are two different types of challenges in dealing with the
PLC effect. First, understanding the collective behavior of
dislocations which has been slow largely due to lack of techniques
for describing the cooperative behavior of dislocations. Second,
the PLC effect involves collective modes of dislocations where
both fast and slow times scales play equally important role which
requires specific techniques of nonlinear dynamics (as we shall
see). Further, these time scales themselves evolve as a function
of strain rate and temperature which in turn leads to different
types of serrations. At low strain rate, the existence of both
fast ( time scales over which stress drops occur) and slow time
scales ( loading time scales) are clearly displayed in the stress
-strain curves. However, at high strain rate, as internal
(plastic) relaxation is not complete, and a clear demarkation of
time scales becomes difficult. This along with the corresponding
length scales ( band widths), which also evolve, points to an
extremely complex underlying dynamics.

The inherent nonlinearity and the presence of multiple time scales
necessitates the use of tools and concepts of nonlinear dynamics
for a proper understanding of this phenomenon. Early theories
based on DSA do not deal with the temporal aspect
\cite{van,penning,Kubin85} and thus are unsuitable for analyzing
the dynamical aspects of the PLC effect. The first dynamical
approach was undertaken in early 80s by Ananthakrishna and
coworkers \cite{Anan82}, which by its very nature affords a
natural basis for the description of the time dependent aspects of
the PLC effect. Further, it also allows for explicit inclusion and
interplay of different time scales inherent in the dynamics of
dislocations. The original model which attempts to address  the
time dependence of the phenomenon uses three types of dislocation
densities assumed to represent the collective degrees of freedom
of dislocations \cite{Anan82}. Despite the simplicity of the
model, many generic features of the PLC effect such as the
existence of a window of strain rates and temperatures within
which it occurs, etc., were correctly reproduced. More
importantly, the {\it negative SRS was shown to emerge naturally}
in the model, as a result of nonlinear interaction of the
participating defects \cite{Anan82,Rajesh}.

Due to the dynamical nature of the model, one prediction that is
unique to this model is the existence of the chaotic stress drops
in a certain range of temperatures and strain rates \cite{Anan83}.
This triggered a series of experiments to verify this prediction.
The method  followed was to analyze the stress-time series
\cite{Anan95,Noro97} using dynamical methods \cite{Licht,Abar}.
Apart from confirming the chaotic nature of stress drops in a
window of strain rates,  these attempts have shown that a wealth
of dynamical information can be extracted from the stress-time
series obtained during the PLC effect \cite{Anan95,Noro97}.
Indeed, the number of degrees of freedom estimated from the
experimental time series turn out to be same as in the model
offering justification for ignoring spatial degrees of freedom.
Subsequent efforts to extend this analysis to the time series
obtained over a range of strain rates showed {\it an intriguing
crossover from a chaotic state at low and medium strain rates to a
power law state at high strain rates} \cite{Anan99,Bhar01}. As the
crossover is observed in both single and polycrystals, it  appears
to be insensitive to the microstructure. However, chaotic state is
dynamically a distinct state from the power law state as the
former involves a small number of degrees of freedom characterized
by the self-similarity of the attractor and sensitivity to initial
conditions \cite{Abar} while latter is an infinite dimensional
state reminiscent of self-organized criticality (SOC)
\cite{Bak,Bak96,Jensen}. Due to this basic difference in the
nature of the dynamics, most systems exhibit either of these
states. More importantly, these studies also demonstrate that the
nature of the dynamics in a given strain regime is correlated with
the nature of band type. The chaotic state has been identified
with the type B bands and the scaling regime at high strain rate
with the propagating type A bands \cite{Bhar01}. These authors
also make a connection between the transition in the nature of
serration between  the type B and type A bands regime of strain
rates with the Anderson's transition in condensed matter physics.
Indeed, recently the spatio-temporal features of the PLC effect
have attracted attention from physicists also \cite{Dann}. Thus,
it appears that the PLC effect is a storehouse of many paradigms
in condensed matter physics. Understanding these connections
between dynamics and general features of   the PLC effect would
give insight into the rich physics. As the above studies
underscore the importance of nonlinearity, it  demands a dynamical
approach to the PLC effect.

The dynamics of the crossover as a function of strain rate is {\it
unusual in a number of ways}. First, the PLC effect is one of the
two rare instances where such an intriguing crossover phenomenon
is seen, the other being in the hydrodynamic turbulence
\cite{Lib}. Second, the power law, both in the PLC effect and
turbulence, arises at high drive rates \cite{Lib,Bohr}. Thus, it
would be interesting to examine the similarity and differences
with hydrodynamic turbulence by comparing results of the Lyapunov
spectrum of the model for the PLC effect and GOY shell model of
turbulence \cite{Yam87,Bohr}.  Another motivation is that such a
study helps us to compare the nature of the Lyapunov spectrum with
the conventional SOC systems seen at low drives ( such as those in
earthquakes \cite{Guten}, acoustic emission during volcanic
activity \cite{Diod}, Barkhausen noise \cite{Barkhou}).   ( For
lack of anything better, we shall reserve SOC for power law
situations at low drives.) Finally, as different types of bands
are a characteristic feature of the PLC effect, we investigate the
connection between spatial aspects and the nature of the dynamics.

The fully dynamical nature of the model and it prediction of
chaotic stress drops  at intermediate strain rates as found in
experiments, makes it most suitable  for studying this crossover
by including  spatial degrees of freedom. The paper reports a
detailed investigation of all these issues ( some of which has
been reported in brief earlier \cite{Bhar02,Bhar03}) within the
context of an extension of Ananthakrishna's model for the PLC
effect. Particular attention will be paid to study the system size
effects during the crossover.

Section II, briefly introduces the dynamical model and its
extension  to include spatial degrees of freedom. Section III
contains the numerical procedure used.  In Section IV, we
introduces the background material  used for the study. Section V
contains  a comparison of the results of analysis of experimental
time series with that of the model. Section VI, contains all the
major results on the dynamics of crossover including the evolution
of the Lyapunov spectrum with special attention to study the
system size effects as a function of the strain rate along with
the analysis of the distribution of null exponents in the power
law regime of stress drops. The section also includes a comparison
of the results of the model with that of the GOY model for
turbulence followed by the slow manifold method of visual
realization of dislocation configurations. Finally, in section VII
we discuss both analytical and numerical results on the nature of
dislocation bands.  We conclude the paper with a few general
comments.

\section{\bf THE ANANTHAKRSHANA's MODEL}

In the dynamical model due to Ananthakrishna and coworkers
\cite{Anan82}, the well separated time scales mentioned in the DSA
are mimicked by three types of dislocations, namely, the fast
mobile, immobile and the 'decorated' Cottrell type dislocations.
The basic idea of the model is that all the qualitative features
of the PLC effect emerge from the nonlinear interaction of these
few dislocation populations, assumed to represent the collective
degrees of freedom of the system. As the model has been studied in
detail by our group and others including an extension to the case
of fatigue \cite{Bekele,Glazov,Zaiser}, following the notation in
Ref. \cite{Rajesh}, we shall briefly outline the model in the
scaled variables. In our model, a natural basis for including the
spatial coupling is through the cross-slip mechanism  proposed
earlier \cite{KFA} with an important difference (see below). The
model consists of densities of mobile, immobile, and Cottrell's
type dislocations denoted by $\rho_m(x,t)$, $\rho_{im}(x,t)$ and
$\rho_c(x,t)$ respectively, in the scaled form.  The evolution
equations are:
\begin{eqnarray}
\nonumber
 \frac{\partial{\rho_m}}{\partial t} & = & -b_0\rho_m^2
-\rho_m\rho_{im} +\rho_{im} - a \rho_m + \phi_{eff}^m\rho_m \\
&+&\frac{D}{\rho_{im}}\frac{\partial^2 (\phi_{eff}^m(x)\rho_m)}{\partial x^2},\label{Eq: xeqn}\\
  \frac{\partial{\rho_{im}}}{\partial t} & = & b_0(b_0\rho_m^2
-\rho_m\rho_{im} -\rho_{im}+a\rho_c), \label{Eq: yeqn}\\
\frac{\partial{\rho_c}}{\partial t} & = & c(\rho_m-\rho_c)
\label{Eq: zeqn}.
\end{eqnarray}
\noindent The model includes the following dislocation mechanisms:
immobilization of two mobile dislocations due to the formation of
locks ($b_0 \rho_m^2$), the annihilation of a mobile dislocation
with an immobile one ($\rho_m\rho_{im}$),  the remobilisation of
the immobile dislocation due to stress or thermal activation
($\rho_{im}$). It also includes the immobilisation of mobile
dislocations due to solute atoms ($a\rho_m$). Once a mobile
dislocation starts acquiring solute atoms we regard it as the
Cottrell's type dislocation $ \rho_c$. As they progressively
acquire more solute atoms, they eventually stop, then they are
considered as immobile dislocations $\rho_{im}$. Alternately, the
aggregation of solute atoms can be regarded as the definition of
$\rho_c$, ie., $\rho_c = \int_{-\infty}^t
dt^{\prime}\rho_m(t^{\prime} ) K(t-t^{\prime})$, where $K(t)$ is
an appropriate kernel. For the sake of simplicity, this kernel is
modelled through a single time scale, $K(t) = e^{-ct} $ . The
convoluted nature of the integral physically implies that the
mobile dislocations to which solute atoms aggregate earlier will
be aged more than those which acquire solute atoms later (see ref.
\cite{Rajesh}). The fifth term in Eqn.(1) represents the rate of
multiplication of dislocations due to cross-slip. This depends on
the velocity of the mobile dislocations taken to be $ V_m(\phi) =
\phi_{eff}^m$, where $\phi_{eff} = (\phi - h \rho_{im}^{1/2})$ is
the scaled effective stress, $\phi$ the scaled stress, $m$ the
velocity exponent and $h$ a work hardening parameter.

The nature of the spatial coupling in the PLC effect has been a
matter of much debate \cite{KFA}. Several mechanisms have been
suggested as a source of spatial coupling, such as compatibility
stresses between the slipped and the un-slipped regions,  long
range interactions, and triaxiality of stresses \cite{KFA}. Within
the scope of our model, cross-slip is a natural source of spatial
coupling, as dislocations generated due to cross slip at a point
spread over to the neighboring elements. Let $\Delta x$ be an
elementary length. Then, the flux $\Phi(x)$ flowing from $x \pm
\Delta x$ and out of $x$ is given by
\begin{equation}
\Phi(x) + \frac{p}{2} \left[\Phi(x+\Delta x) - 2 \Phi(x) + \Phi(x
- \Delta x)\right] ).
\end{equation}
where $\Phi(x) = \rho_m(x)V_m(x)$ and $p$ is the probability of
cross-slip spreading into neighboring elements. Expanding $\Phi(x
\pm \Delta x)$ and keeping the leading terms, we get
\begin{equation}
\rho_m V_m + \frac{p}{2}\frac{\partial^2(\rho_m V_m)}{\partial
x^2} (\Delta x)^2.
\end{equation} We further note that cross-slip
spreads only into regions of minimum back stress. Here, we
consider the back stress is taken to result from the immobile
dislocation density ahead of it. Thus, we use $\Delta x^2 =
<\Delta x^2> = \bar {r}^2 \rho_{ im}^{-1}$, where  $<\ldots>$
refers to the ensemble average and $\bar{r}^2$ is an elementary
(dimensionless) length. Finally, $a$, $b_0$ and $c$ are the scaled
rate constants referring, respectively, to the concentration of
solute atoms slowing down the mobile dislocations, the thermal and
athermal reactivation of immobile dislocations, and the rate at
which the solute atoms are gathering around the mobile
dislocations. We note here that the order of magnitudes of the
constants have been identified in Ref.
\cite{Anan82,Bekele,Zaiser}. These equations are coupled to the
machine equation
\begin{equation}
\frac{d\phi(t)}{dt}=
d[\dot{\epsilon}-\frac{1}{l}\int_0^l\rho_m(x,t)
\phi_{eff}^m(x,t)dx], \label{Eq: seqn}
\end{equation}
where $\dot\epsilon$ is the scaled applied strain rate, $d$ the
scaled effective modulus of the machine and the sample, and $l$
the dimensionless length of the sample. (We reserve $\dot
{\epsilon}_a $ for the unscaled strain rate.) We also note here
that there is a feed back mechanism between Eq. \ref{Eq: seqn} and
Eq.(1). The machine equation which determines the stress depends
on the the difference between the applied strain rate and average
plastic strain rate generated in the sample. Thus, the nature of
internal relaxation can influence stress generated in the sample
which in turn determines the dislocation multiplication in Eq.
(1). This type of global coupling (Eq. \ref{Eq: seqn}) is common
to many other situations for instance in the nonlinear transport
properties of charge density waves ( in blue bronze for example)
\cite{Dumas}. We shall make some comments on this later.

\section{NUMERICAL SOLUTION OF THE MODEL}

We first note that the spatial dependence of $\rho_{im}$ and
$\rho_c$ arises only through that of $\rho_m$. We solve the above
set of equations by discretizing the specimen length into $N$
equal parts. Then, $\rho_m(j,t)$, $\rho_{im}(j,t)$, $\rho_c(j,t)$,
$j= 1,...,N$, and ${\phi}(t)$ are solved. The widely differing
time scales \cite{Rajesh,Rajesh00,Bhar02} calls for appropriate
care in the numerical solutions. We use a variable step fourth
order Runge-Kutta scheme with an accuracy of $10^{-6}$ for all the
four variables. The spatial derivative in $\rho_m$ is approximated
by its central difference. The initial values of the dislocation
densities are so chosen that they mimic the values in real
samples. They are uniformly distributed with a Gaussian spread
along the sample. However, for most calculations, we have used the
steady state values for the variables as the long term evolution
does not depend on the initial values. As for the boundary
conditions, we note that the sample is strained at the grips. This
means that there is a high density of immobile dislocations at the
ends of the sample. We simulate this by employing two orders of
magnitude higher values for $\rho_{im}(j,t)$ at the end points $j
= 1$, and $N$ than the rest of the sample. Further, as bands
cannot propagate into the grips, we use $\rho_m(j,t)
=\rho_c(j,t)=0$ at $j=1$  and $N$.

\begin{figure}
\includegraphics[height=4.5cm,width=7.5cm]{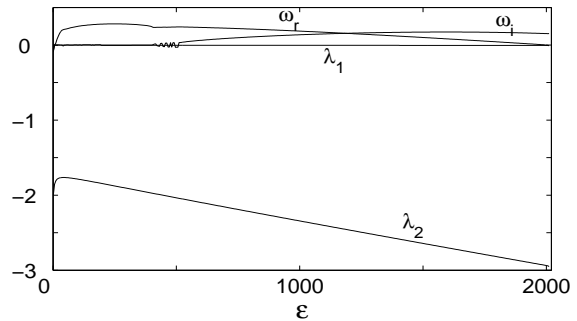}
\caption{Eigen value spectrum of the fixed point for the model.
$\omega_r$ and $\omega_i$ refer to the real and imaginary parts of
the eigen value.}
 \label{fig0}
\end{figure}

For the original model ($D =0$), it has been shown that the fixed
point of the system of equations becomes unstable in a certain
range of parameter values. In particular, as a function of the
applied strain rate, the PLC state is reached through a Hopf
bifurcation and is terminated by a reverse Hopf bifurcation ( with
the other parameters kept in the instability domain). This feature
is retained with the addition of the spatial degrees of freedom
except that the number of complex conjugate roots are 2N, the
negative ones are N and one zero exponent. We find that the
instability domain in $\dot\epsilon$ increases when the values of
the other parameters $a,b_0,c,d,m$ are taken as in Ref.
\cite{Rajesh00}. This is due to fact that the range of $\dot
\epsilon$ depends on the value of $D$ due to the global coupling
in Eq. \ref{Eq: seqn}. ( The domain converges quickly as a
function of $N$.) The boundary of $ \dot \epsilon $ is
approximately in the range 10 to 1000 for $a = 0.8, b_0 = 0.0005,
c = 0.08, d =0.00006, m = 3.0$, $h=0$ with $D = 0.5$, beyond which
a uniform steady state exits. A set of four eigen values are shown
in Fig. \ref{fig0}. The numerical results reported in the present
work are for the above values. However, the results hold true for
a wide range of values of other parameters in the instability
domain including a range of values of $D$. Various system sizes
are used depending upon the property studied, but  are  generally
in the range $N$=100 to 3333. A sequence of values of N are used
wherever convergence of the properties are investigated.

\section{METHODOLOGY}

As our approach is fully dynamical and keeping in view the
materials science community, we collect here a few definitions and
provide some details of the methodology used in the analysis.
Characterizing the dynamics of the model equations is carried out
by studying the Lyapunov spectrum.  The number of Lyapunov
exponents $M$ for a given $N$ is $M=3 N +1$. We shall also use two
other well know invariants namely the Kaplan-Yorke dimension
$D_{KY} =j+\frac{\sum^j_{i=1}\lambda_i}{\vert\lambda_{j+1}\vert}$,
where $j$ is such that  $\sum^{j}_{i=1}\lambda_i>0,
\sum^{j+1}_{i=1}\lambda_i <0$ and the Kolmogorov entropy $ H
=\sum^p_{i=1}\lambda_i$ such that $\lambda_p\ge 0 $ and
$\lambda_{p+1} <0$.  One important issue that is relevant to
systems with many degrees of freedom is the existence of a
limiting density for the Lyapunov spectrum as the system size is
increased.  This requires that we should ascertain if $\lambda_j$
versus $x=j/L^d$ converges to a well-defined asymptotic density
function $\Lambda(x)$ with $x \in [0,1]$. ( See Ref. \cite{Bohr}.)
We address this issue by calculating the spectrum for various
system sizes $N$ = 100 to 3333 which covers approximately two
orders in $M$. In particular, such a study will be useful in
comparing the results of our model with the GOY shell model for
turbulence \cite{Yam87} in the power law regime of stress drops.
Then, one expects that $j/D_{KY}$ converges to well defined
density function. Following Ref. \cite{Yam87}, we use $j/D_{KY}$
verses an appropriately scaled quantity $\lambda_j D_{KY}/H$. This
quantity is expected to converge to $f( \lambda_j D_{KY}/H$). ( We
note here that the distribution function is proportional to the
negative derivative  of $f$.) The nature of the converged Lyapunov
density function $f(\lambda_j D_{KY}/H)$ as a function of the
drive parameter $\dot \epsilon$ can be used to quantify the
changes the dynamics during the crossover.

As stated earlier, a proper description of the PLC effect requires
a description of  both the slow and fast time scales which in turn
requires special techniques in nonlinear dynamics. These two time
scales are transparent in the model equations where Eq. (\ref{Eq:
xeqn}) represents a fast dynamics compared to the rest ( both Eq.
(\ref{Eq: yeqn}) and (\ref{Eq: seqn}) are slow while   (\ref{Eq:
zeqn}) falls in between). Such a system can be studied by
eliminating the fast variable thereby allowing a reduction in the
dimensionality of the system \cite{Milik}. To illustrate this
consider
\begin{eqnarray}
 \mu  \dot x & = & f(x,y,\mu) \label{fast}\\
  \dot y     & = & g(x,y,\mu)\label{slow}
\end{eqnarray}
where $\mu$ is small parameter and  $x \in \mathbb{R}^p$ and $y
\in \mathbb{R}^q$. The main feature of such systems is that $x$
evolves much faster than $y$ unless $f(x,y,\mu)$ is small. In the
vicinity of the slow manifold defined by $f(x,y,\mu)=0$, the
dynamics is characterized by the evolution of the slow variable
$y$. Thus, there is a reduction in the dimensionality of the
system. On the other hand, if one is interested in the fast
sub-system, using a scaled time $\tau = t/\mu$, we get the
corresponding fast variable $x$ defined by Eq. \ref{fast} where
the slow variables $y$ act as parameters (obtained from Eq.
\ref{slow}). This subspace is clearly the complimentary subspace
of the slow manifold. We shall use these two subspaces for the
visualization of dislocation configurations in the high strain
rate power law regime and obtain the band velocity at high strain
rates respectively.

The analysis of the  experimental stress-time series is carried
out by estimating  both correlation dimension $\nu$ and the
Lyapunov spectrum. These methods involve embedding the scalar time
series in a higher dimensional space using time-delay technique
\cite{Pack,Abar}. Given a time series \{$\sigma_j \vert j=
1,..,M\}$, one first constructs vectors $\vec{\xi}_i = (\sigma_i,
\sigma_{i - \tau}, \sigma_{i - 2\tau},...,\sigma_{M - (d-1)\tau})$
in a $d$ dimensional space. The assumption here is that the actual
dynamics can be unfolded by embedding the time series in a higher
dimensional space in which the original attractor resides. (In
addition, surrogate data analysis was also carried out in
\cite{Anan99}.) Then, a quantitative estimate of the self
similarity of the attractor, namely the correlation dimension,
$\nu$, can be obtained by calculating the integral \cite{Grass}
$C(r) = \frac{1}{N_T} \sum \Theta ( r - \vert \vec {\xi}_i - \vec
{\xi}_j \vert ) \sim r^{\nu}$, where $N_T$ is the total number of
points in the sum. Correlation dimension also provides a lower
bound for the number degrees of freedom required for a dynamical
description of the system which is given by the minimum integer
larger than $\nu +1$ \cite{Ding}. The geometrical interpretation
of these degrees of freedom is that they correspond to the
subspace to which the trajectories are confined.
\begin{figure}
\includegraphics[height=6cm,width=8cm]{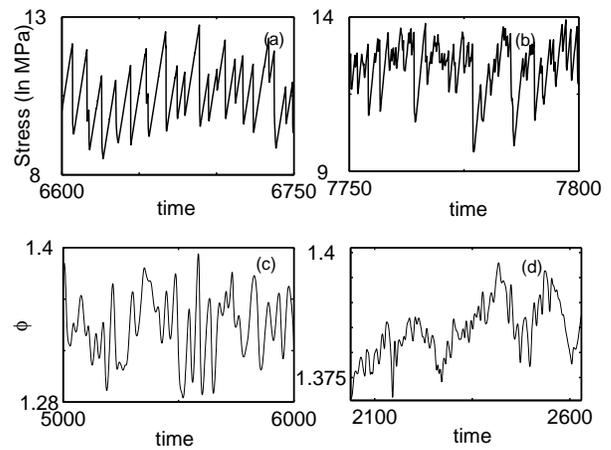}
\caption{(a) \& (b) Experimental stress-time series: (a) chaotic
state at strain rates $\dot\epsilon_a = 1.7\times 10^{-5}s^{-1}$
and (b) power law state at $\dot\epsilon_a=8.3
\times10^{-5}s^{-1}$.  (c) \& (d) Stress-time series from the
model at (c) $\dot\epsilon=120$ (d)$\dot\epsilon=280$.}
\label{fig1}
\end{figure}
The dimension of this subspace can be obtained directly by using
singular value decomposition (SVD) \cite{King}. This method is
often used for filtering noise component superposed on the time
series. However, another use of SVD in the present context is that
it is useful for  the {\it visualization of the strange
attractor}. (This method has been applied to the PLC time series
earlier \cite{Noro97}.) The method involves setting up the $m
\times d$ trajectory matrix $\bf T$ defined by
$(\vec{\xi}_1,\vec{\xi}_2,...,\vec{\xi}_m)$ where $m = M - (d-1)
\times \tau$. The eigen values of the matrix are obtained using
the standard method of decomposition $ \bf T = \bf{U} \bf{W}
\bf{V}^T$, where $\bf U$ is $m \times d$ orthogonal matrix, $\bf
V$ is a $d\times d$ unitary matrix and $\bf W$ is the matrix of
eigen values of the covariance matrix  of $\bf T$ which are all
nonnegative. The eigen values usually decrease rapidly saturating
to a level below which the changes are minimal. Then the dimension
of the attractor is taken to be that corresponding to number at
which the eigen values saturate.

\section{COMPARISON WITH EXPERIMENTS}

To make the motivation clear, we begin by briefly recalling the
relevant experimental results on the crossover phenomenon and then
compare them with those from the model. The simplest feature to
compare is the nature of serrations in the respective regimes of
strain rate. Plots of two experimental stress-strain curves from
$Cu Al$ single crystals corresponding to the chaotic and power law
regimes of applied strain rates are shown in Fig. ~\ref{fig1} a,b.
The stress-time series in the intermediate and high strain rate
regimes from the model are shown in Fig.~\ref{fig1}c,d. The
similarity of the experimental time series with that of the model
in the respective regimes are clear.
\begin{figure}
\includegraphics[height=5cm,width=8.5cm]{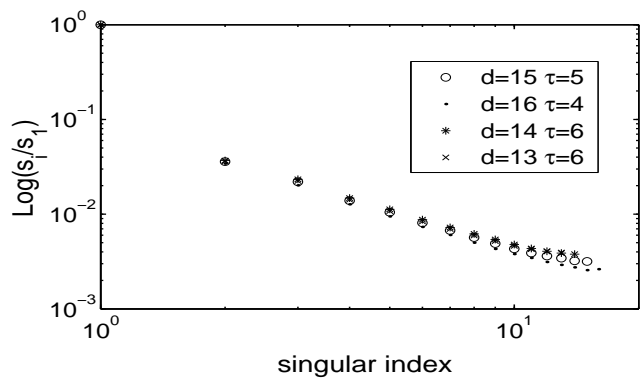}
\caption{Singular value spectrum of the experimental time series
shown in Fig.~\ref{fig1}a}. \label{fig2}
\end{figure}

\begin{figure}
\includegraphics[height=4.5cm,width=6.5cm]{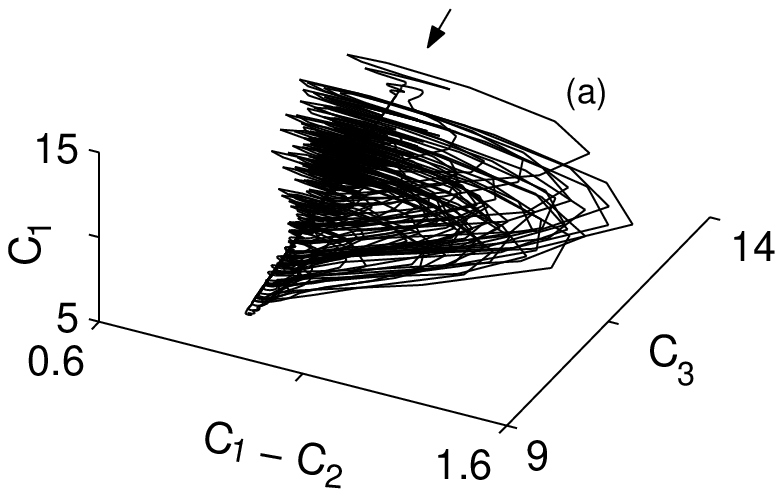}
\includegraphics[height=4.5cm,width=6.5cm]{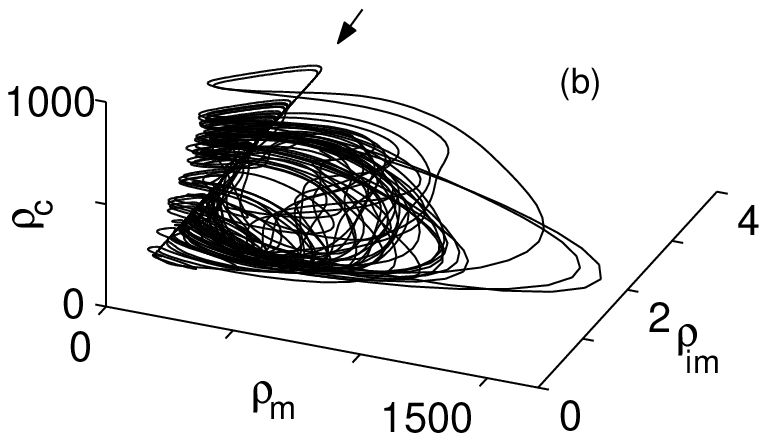}
\caption{(a) Reconstructed experimental attractor from the time
series shown in Fig.~\ref{fig1}a. (b)Attractor from the model for
$N=100$, $j=50$. } \label{fig3}
\end{figure}

The analysis of the stress-time series given in Fig. \ref{fig1}
has been reported in Ref. \cite{Anan99}. The correlation dimension
was found to be  $\nu = 2.3$. Then, the number of degrees of
freedom required for the description of the dynamics of the system
given by the minimum integer larger than $\nu +1$ \cite{Ding} is
seen to be four, consistent with that used in the original model.
As an independent check to obtain the number degrees of freedom,
as also for the visualization of the experimental attractor, we
have carried out the singular value decomposition of this time
series. The normalized eigen values ( with respect the largest) is
shown in Fig. ~\ref{fig2}. It is clear that the relative strength
of the fourth eigen value drops more than two orders of magnitude
compared to the first and changes very little beyond the fourth.
Thus, we estimate the dimension of the experimental attractor to
be four which is also consistent with that obtained from the
correlation dimension. ( For time series from model systems, one
usually finds a floor level below which the eigen values saturate.
This is taken as the dimension of the actual attractor. However,
in real situations, as in the present case, the eigen values do
not saturate due to the presence of noise.) Then, for the
visualization of the experimental attractor, we can use the
dominant eigen values to reconstruct the nature of the attractor.
Using the first three principal directions of the subspace $C_i; i
=1$ to 3, we have reconstructed the experimental attractor in the
space of specifically chosen directions $C_1 - C_2, C_3$ and $C_1$
to permit comparison with the attractor obtained from the model.
This is shown in Fig. ~\ref{fig3} a for the experimental time
series at $\dot {\epsilon}_a = 1.7 \times 10^{-5}s^{-1}$. This can
be compared with the strange attractor obtained from the model in
the space of $\rho_m,\rho_{im}$ and $\rho_c$ (at an arbitrary
spatial location, here $j=50$ and $N=100$) shown in
Fig.~\ref{fig3} b for $\dot \epsilon = 120$ corresponding to the
mid chaotic region (see below).  Note the similarity with the
experimental attractor particularly about the linear portion in
the phase space (Fig.~\ref{fig3} a). This direction can be
identified with the loading direction in Fig.~\ref{fig1} a.  Note
that the identification of the loading direction is consistent
with the absence of growth of $\rho_m$.

In contrast to the experimental time series at low and medium
strain rates, for the time series at the highest strain rate (
Fig. \ref{fig1} b), we neither find a positive Lyapunov exponent
and nor a converged value of the correlation dimension as was
shown in Ref.\cite{Anan99}. However, the distribution of  stress
drops was shown to obey a power law \cite{Anan99}.

\begin{figure}
\includegraphics[height=4.5cm,width=7.5cm]{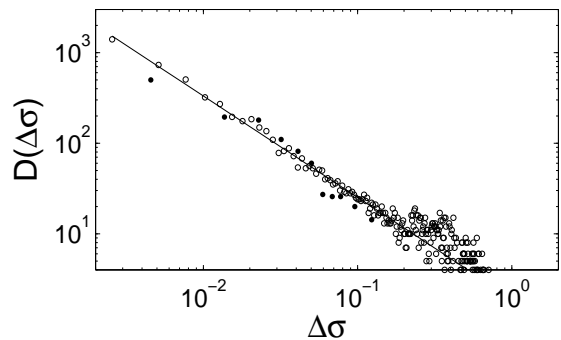}
\caption{Distributions of the stress drops from the model
($\circ$), from experiments ($\bullet$)  for $N = 1000$ and
$\dot\epsilon =280$. Solid line is a guide to the eye.}
\label{fig4}
\end{figure}

Apart from comparing the statistics of stress drops from the model
with that of the experimental time series, there is another more
important issue, namely, does it generate power law statistics? If
so,  what mechanism is responsible for this? This is particularly
important as the model is fully dynamical and noise free. It is
clear that Fig.\ref{fig1} d is similar to Fig. \ref{fig1} b, as
there no inherent scale in the magnitudes of the stress drops in
both cases  and thus it is likely to also show a power law
statistics. Indeed, the distribution of stress drop magnitudes,
$D(\Delta \phi)$, shown in Fig. \ref{fig4} obtained from long runs
for a large system size ( $N =1000$) shows a power law $D(\Delta
\phi) \sim \Delta \phi ^{- \alpha }$ over two orders of magnitude
which increases with both length of stress series and system size.
( Note that the value of $N$ here nearly three times larger than
the results in Ref. \cite{Bhar02}, Fig. 3b and thus, the power law
is well converged with respect to the system size. ) Surprisingly,
experimental points ($\bullet$) corresponding to $\dot
{\epsilon}_a = 8.3 \times 10^{-5} s^{-1}$ also fall on the same
curve with an exponent value $\alpha \approx 1.1$. ( We have
scaled the experimental points by a constant amount along both the
axis to show that these points also fall on the same line.)  The
distribution of the duration's of the stress drops $D(\Delta t)
\sim \Delta t^{ -\beta}$ also shows a power law with an exponent
value $\beta \approx 1.3$. The conditional average of $\Delta \phi
$ denoted by $<\Delta\phi>_c$ for a given value of $\Delta t$
behaves as $<\Delta\phi>_c \sim {\Delta}t^{1/x}$ with $x \approx
0.65$. The exponent values satisfy the scaling relation $\alpha =
x(\beta - 1)+1$ quite well. The exponent values remain unaltered
in the region of strain rate $270 < \dot \epsilon < 700$ we have
investigated thus is independent of the value of the drive
parameter. ( There are models of coupled map lattices that produce
power laws where the exponent value depends on the drive
parameter.) We now investigate  the underlying causes leading to
this power law.

\section{ DYNAMICS OF CROSSOVER}

\subsection{Lyapunov Spectrum}

Our next aim is to characterize the dynamics of this crossover. A
natural tool for characterizing the crossover is to study the
distribution of Lyapunov exponents as a function of the applied
strain rate in the entire interval where the PLC effect is seen.
Further, we also discuss the convergence properties of the
Lyapunov spectrum as the system size is increased. In particular,
this will be useful in examining the density of null ( nearly
vanishing) exponents and also to compare our results with that of
the GOY model of turbulence.

We have calculated the spectrum of Lyapunov exponents using the
algorithm due to  Benettin {\it et al} \cite{Bennet}. The exponent
values reported here were obtained by averaging over 15000 time
steps after stabilization with an accuracy of $10^{-6}$. We have
used several system sizes ranging from $N=100, 150,\, 350, \,500,
1000$, and 3333 which covers approximately two orders of magnitude
in $M$, {\it i.e.,} from 301 to 10000. A rough idea of the changes
in the dynamics of the system can be obtained by studying the
dependence of the largest Lyapunov exponent (LLE) as a function of
the strain rate. The LLE converges fast as a function of the
system size. For instance,  we find that the LLE for $N=500$ looks
much the same for a much smaller system size N=100 given in Fig.
3a in Ref. \cite{Bhar02}.  The LLE becomes positive around $\dot
\epsilon \approx 35$ reaching a maximum at $\dot \epsilon =120$,
practically vanishing around 250. ( Periodic states are observed
in the interval $10 <\dot \epsilon < 35$.) In the region $\dot
\epsilon \ge 250$, the dispersion in the value of the LLE is $
\sim 5 \times 10 ^ {-4}$ which is the same order as the mean.
Thus, the LLE can be  taken to vanish beyond $\dot\epsilon = 250$.

\begin{figure}
\includegraphics[height=5.5cm,width=8cm]{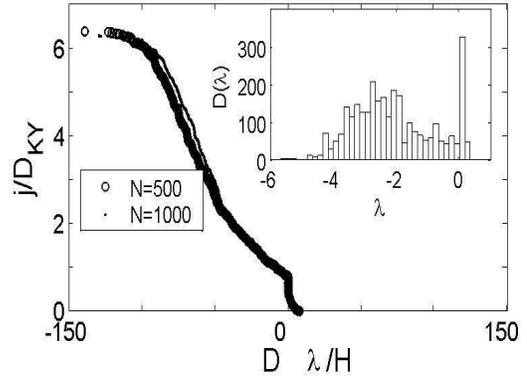}
\caption{ A plot of $j/D_{KY}$ verses $\lambda_j D_{KY}/H$ for
$\dot\epsilon=120$. Inset shows a plot of
$D(\lambda)$ as a function of $\lambda$ for N =1000.} \label{fig5}
\end{figure}

The study of the Lyapunov spectrum  reveals that in the chaotic
regime of strain rates,  only a small proportion of the exponents
are  positive, an equal small number are close to zero value and a
large proportion of the exponents are negative. The distribution
of the Lyapunov exponents $D(\lambda)$, is shown for $N=1000$ in
the inset of Fig. ~\ref{fig5} for the strain rate,
$\dot\epsilon=120$. For this system size (with a total of
exponents $M$=3001), the number of positive exponents is $\approx$
6.2\% of the total number of exponents and the null exponents  are
also $\approx$ 9\%. (For numerical purposes null exponents are
taken to correspond to $|\lambda|\le5.2\times10^{-4}$.) These
ratios remain the same for the larger system sizes used.

\begin{figure}
\includegraphics[height=5cm,width=8cm]{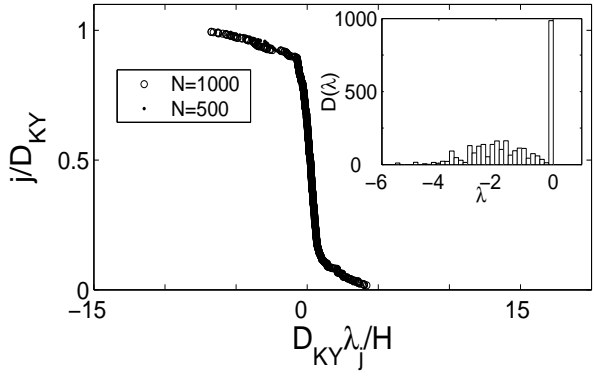}
\caption{A plot of $j/D_{KY}$ verses $\lambda_j D_{KY}/H$ for
$\dot\epsilon=220$. Inset shows a plot of
$D(\lambda)$ as a function of $\lambda$ for N =1000. }
\label{fig6}
\end{figure}

\begin{figure}
\includegraphics[height=5cm,width=8cm]{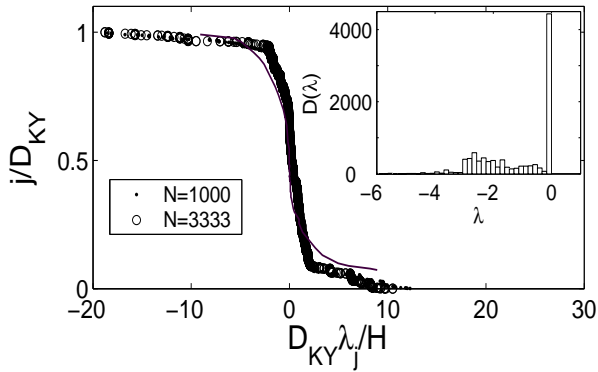}
\caption{A plot of $j/D_{KY}$ verses $\lambda_j D_{KY}/H$ for
$\dot\epsilon=280$. Inset
shows a plot of $D(\lambda)$  for N =
3333. A schematic plot of the Lyapunov density function
(continuous line) for the GOY model (after \cite {Yam87}).}
\label{fig7}
\end{figure}

While $D(\lambda)$ reflects the distribution of Lyapunov exponents
in various regions, for studying the convergence of the Lyapunov
spectrum, plots of   the density function $j/D_{KY} = f( \lambda_j
D_{KY}/H)$ are better suited. Further, these quantities have been
used traditionally in the studies of extended dynamical systems
\cite{Bohr}.  A plot of $j/D_{KY}$ verses $\lambda_j D_{KY}/H$ for
$\dot \epsilon = 120$ for N = 500 and 1000 is shown in Fig.
\ref{fig5}. It is clear that while the density function has not
yet converged for negative values of $\lambda_j D_{KY}/H$, those
for positive values are already converged. As we increase the
strain rate beyond $\dot\epsilon=180$, concomitant with the
decrease in the value of the LLE, the number of null exponents
increases. For instance, at $\dot\epsilon=220$, for which the
maximum Lyapunov exponent is small $\sim$ 0.0058, the number of
null exponents increases  to 30\% $M$ (see inset of Fig.
\ref{fig6}).  $D(\lambda)$ shows that the number of null exponents
has increased. Concomitant with this trend,   a plot of $j/D_{KY}$
verses $\lambda_j D_{KY}/H$ for N = 500 and 1000 (Fig. \ref{fig6})
shows that for $\dot \epsilon =220$ is well converged for the
entire range of values of the scaled Lyapunov exponent $ \lambda_j
D_{KY}/H$. This signals a faster convergence of the density
function $j/D_{KY} = f( \lambda_j D_{KY}/H)$  with the system size
as we approach the scaling regime. Indeed, we find that
plots for N = 500 and 1000 for strain rate $\dot \epsilon =280$
cannot be distinguished over the entire range of values of
$\lambda_j D_{KY}/H$.  Even though it would be adequate to use
$N=1000$, for further analysis when dealing with Lyapunov spectrum
in  the scaling regime, we use a much bigger system size of
$N=3333$, which for all practical purposes can be taken to be
large $N$ limit. A plot of $j/D_{KY} $ verses $\lambda_j D_{KY}/H$
shown in Fig. \ref{fig7} for $N$ = 3333 ( and also for 1000) shows
that for $\dot \epsilon =280$ is well converged for the entire
range of values of $ \lambda_j D_{KY}/H$. Note also that nearly
40\% the exponents are close to zero (see the inset).

\begin{figure}
\includegraphics[height=5cm,width=8cm]{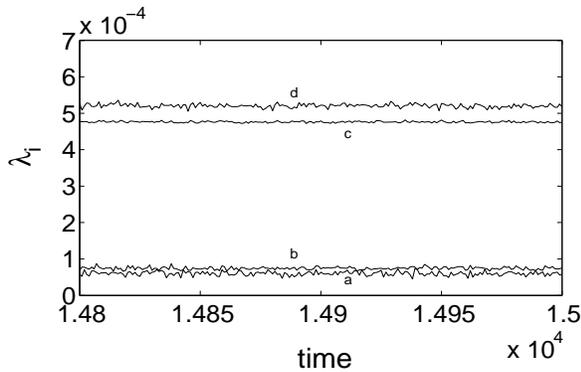}
\caption{The first two Lyapunov exponents  that do not cross each
other as a function of time for N =3333 for $\dot\epsilon$=280.
Also shown are two more exponents that are close to each other.}
\label{fig8}
\end{figure}

\begin{figure}
\includegraphics[height=5cm,width=8cm]{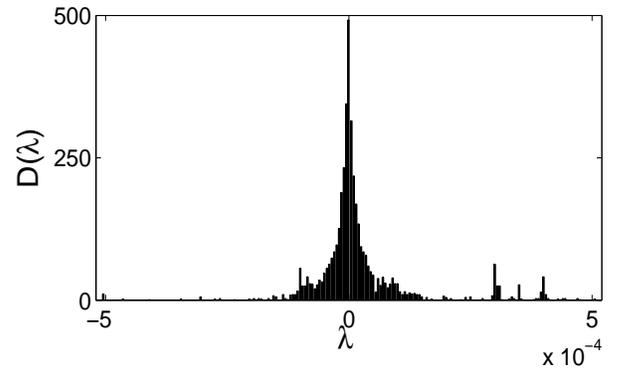}
\caption{The peaked nature of the distribution of null exponents
lying in the range $[-5.2\times 10^{-4}, 5.2 \times 10^{-4}]$ for
$\dot\epsilon=280, N=3333$.} \label{fig9}
\end{figure}

As we approach the power-law regime of stress drops (extending
from $\dot\epsilon=250$), as the largest Lyapunov exponent
approaches  zero ($\sim 5.16 \times 10^{-4}$ for $\dot \epsilon
=280$), exponents below a certain value cross each other as a
function of time. However, the first few exponents remain
distinct. Figure \ref{fig8} shows the first two exponents that are
well separated and another two which are close to each other in
magnitude (for $\dot \epsilon = 280$ and N = 3333). Below $\vert
\lambda \vert < 5 \times 10^{-5}$, even though the exponents cross
each other, the distribution of the exponents remains unchanged.
The most significant feature of the spectrum in the region is that
there is a {\it dense set of null exponents}.  The peaked nature
of the distribution of the null exponents ($|\lambda| \le
5.2\times10^{-4}$) for $\dot\epsilon=280$ for $N=3333$ is shown in
Fig. ~\ref{fig9}.


\begin{figure}
\includegraphics[height=5cm,width=8cm]{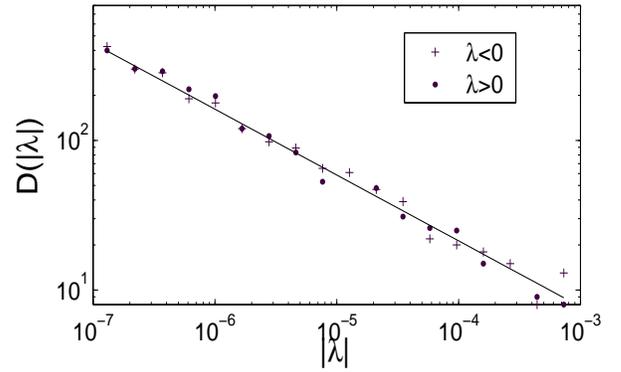}
\caption{Log-log plot of the distribution of the marginal
exponents for $\dot\epsilon=280, \, N=3333$. Solid line is a guide
to the eye.} \label{fig10}
\end{figure}

The peaked nature of $D(\vert \lambda\vert)$  for null exponents
suggests the possibility of a power law distribution for their
magnitudes. We have plotted the distribution of the null exponents
($\vert\lambda_i\vert \le 5.2\times 10^{-4}$) for
$\dot\epsilon=280$, for a system size of $N=3333, \, M=10000$
shown in Fig.~\ref{fig10}.  It is clear that  both the positive
and negative exponents show a power-law distribution $D(|\lambda|)
\sim |\lambda|^{-\gamma}$ with an exponent value $\gamma \sim
0.51$ and the scaling extends over an impressive three decades as
shown in Fig.~\ref{fig10}. As null exponents correspond to
marginal stable nature of the system, their finite density, which
itself obeys a power law, elucidates the underlying cause of power
law distribution of stress drops at high strain rates.

\subsection{Comparison with Shell Model for Turbulence}

As mentioned in the introduction, both in the PLC effect and in
turbulence,  power law statistics is seen at high drive rates in
contrast to conventional SOC systems where the power law arises at
low drives \cite{Jensen}. In addition, the dense set of null
Lyapunov exponents in the scaling regime is similar to the finite
density of null exponents obtained by Ruelle \cite{Ruelle} for the
discrete spectrum of the operator that linearizes the
Navier-Stokes equations, if the fractal dimension of the energy
dissipation set is $D_F \le 2.5$. This property is preserved by
the GOY shell model \cite{Yam87}. Here we attempt a comparison of
the Lyapunov spectrum obtained from our model with that of the GOY
model.

\begin{figure}
\includegraphics[height=5cm,width=8cm]{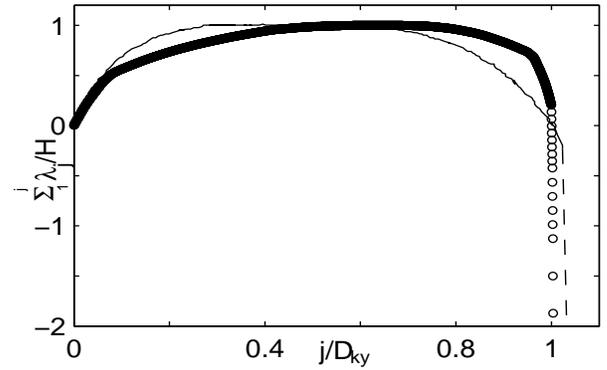}
\caption{A plot of $\sum_1^j\lambda_j/H$ as a function of
$j/D_{KY}$ for  $\dot\epsilon=$ 280 and $N =3333$. The
corresponding schematic plot for the GOY model for $j/D_{KY} <1$
(continuous line) and $j/D_{KY}\ge 1$ (dashed curve) (after \cite
{Yam87}). } \label{fig11}
\end{figure}

Shell models of turbulence \cite{Bohr} are designed to mimic the
behavior of Navier-Stokes equations at high drives where the power
law is seen. One standard model is the GOY model
\cite{Yam87,Bohr}. For this model, Ohkitani and Yamada
\cite{Yam87} gave a good numerical evidence that the density
function exists as the viscosity parameter $\eta$ tends to zero.
In our case, the role of the viscosity parameter is taken by the
applied strain rate. In Section IV A, we have shown that there is
a rapid increase in the density of null exponents and
consequently, there is a rapid convergence of $j/D_{KY} = f(
\lambda_j D_{KY}/H)$ as function of N, starting from  $\dot
\epsilon =220$. This suggests that one should expect convergence
of the limiting $j/D_{KY} = f( \lambda_j D_{KY}/H)$ function as we
approach the power law strain rate regime of  stress drops.  Thus,
we should expect that the limiting distribution itself converges
as a function of $\dot \epsilon$ as we approach the scaling
regime. Considering $N =1000$ approximates the limiting
distribution ( see Fig. \ref{fig6} for justification), we have
verified that plots of $j/D_{KY} = f( \lambda_j D_{KY}/H)$ for
three values of $\dot \epsilon=250, 260$ and 280 for reasonably
large N = 1000 converge. This result is similar to the convergence
of the density function in the GOY model as a function the
viscosity parameter. The density function $j/D_{KY}$ obtained from
the model can be compared with that of the  GOY model. Plot of
$j/D_{KY} = f( \lambda_j D_{KY}/H)$ for a large system N=3333 (
which can be taken to represent the limiting density as a function
of system size ) for $\dot \epsilon =280$ is shown in Fig.
\ref{fig7} along with a schematic plot for the GOY model shown by
the continuous line.  As can be seen, in both the cases, the
distribution function which is proportional to
$-df(\lambda)/d\lambda$, shows a singularity near zero. The
difference being that the singularity is more pronounced for our
model. Ohkitani and Yamada also plot another quantity which
represents the null exponents better, namely, the sum of Lyapunov
exponents up to $j$ normalized by $H$ as function of $j$ scaled by
$D_{KY}$. The quantity $\sum_1^j\lambda_j/H$  is an increasing
function of $j/D_{KY}$ for positive $\lambda_j$ and goes to unity
when $\sum_{i=1}^j\lambda_i =H$. In the region of null exponents,
this quantity remains constant and then decreases with $j$ when
$\lambda_j$'s are negative. Thus, this quantity also reflects the
density of null exponents. A schematic plot of
$\sum_1^j\lambda_j/H$ as a function of $j/D_{KY}$  ( continuous
and dashed line) for the GOY model is shown in Fig. \ref{fig11}.
The increase in $\sum_1^j\lambda_j/H$ for small $j/D_{KY}$ shows
that there is a finite density of positive exponents in the
Lyapunov spectrum for the GOY model. Further, these authors find
that there is a convergence with respect to the viscosity
parameter for Lyapunov spectrum corresponding to the interior of
the attractor (ie., $j/D_{KY} < 1$), while there is scatter for
$j/D_{KY} > 1$ (the dashed line represents this portion).  We have
plotted $\sum_1^j\lambda_j/H$ as a function of $j/D_{KY}$  for N
=3333 on the same plot for the sake of comparison. In our case,
the increase to unit value is much slower ( compared to the GOY
model) which clearly implies that there are very few positive
exponents (of any significant magnitude) with most of them being
vanishingly small. This feature is unlike the shell model where
there is a finite density of positive exponents. In the GOY model,
the largest exponent is proportional to $\eta^{-1/2}$ which is
reflected in the steeper increase in $\sum_1^j\lambda_j/H$ for the
GOY model.

Thus, we have shown that the power-law regime seen in our model at
high drives as in hydrodynamics is not a superficial feature. The
Lyapunov spectrum of the model is quite similar to the GOY model
which however  has a finite density of positive exponents. In
addition, the distribution of the {\it null Lyapunov exponents}
itself shows a power law in our case. Such a feature has not been
verified for the GOY model.  This feature is also quite distinct
from the Lyapunov spectrum of models of SOC studied so far
\cite{Erzan,desouza,Cess}. We comment on this later.

\subsection{Slow Manifold Analysis}

The analysis provided in the previous section shows that as the
strain rate is increased most exponents  get concentrated in a
narrow interval around the zero value. As zero Lyapunov exponents
represent a marginal situation, we see that  the region of strain
rate beyond 240  (corresponding to the power law statistics of
stress drops) can be identified with a marginally stable state.
Thus,  it would be interesting to realize a geometrical picture of
dislocation configurations in the marginal state and examine how
dislocations reach such state as a function of strain rate. We
accomplish this through the slow manifold of the model
\cite{Rajesh00,Rajesh,Milik}.

\begin{figure}
\includegraphics[height=5.0cm,width=8.0cm]{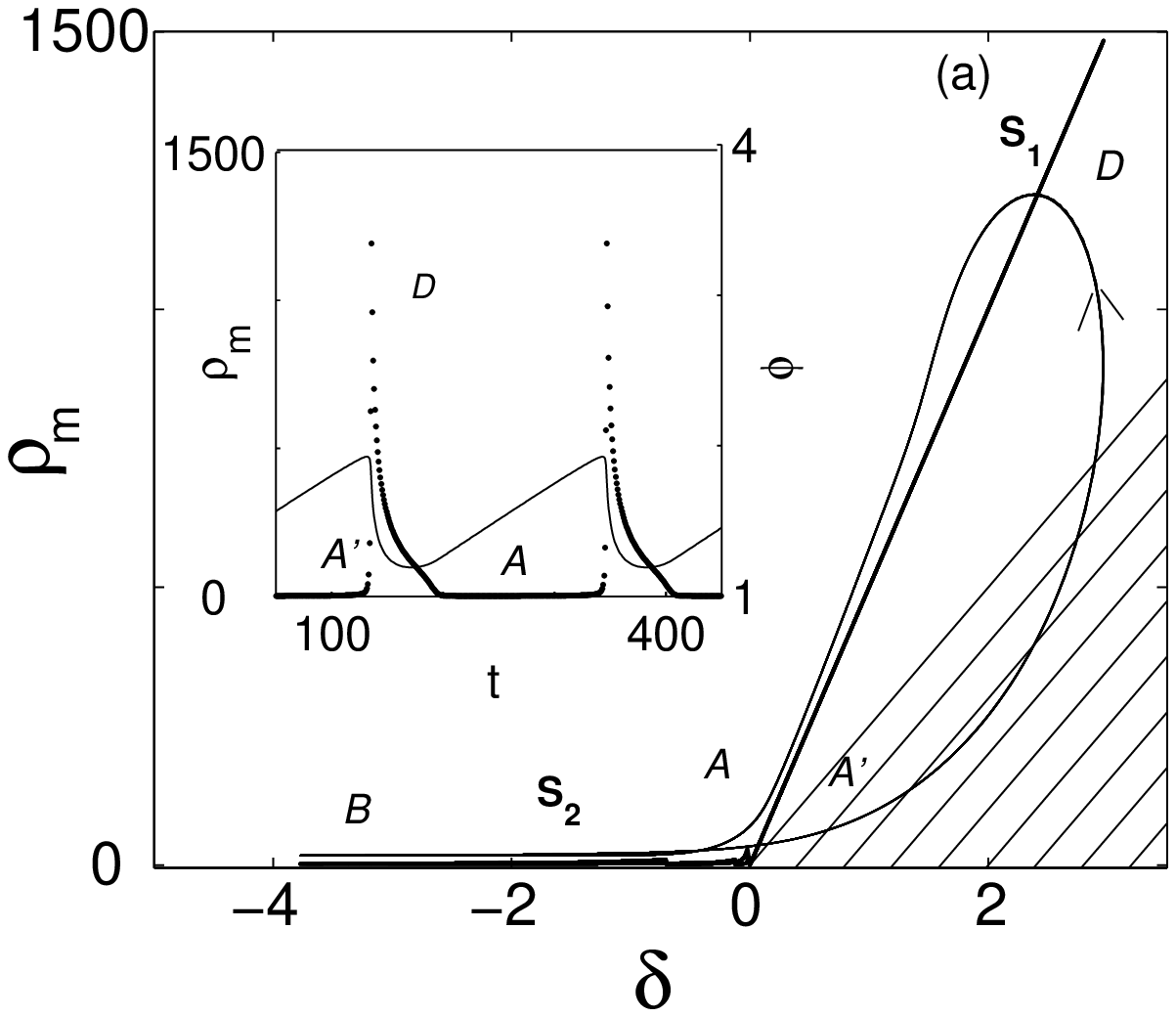}
\vspace{0.5cm}
\includegraphics[height=5.5cm,width=8.0cm]{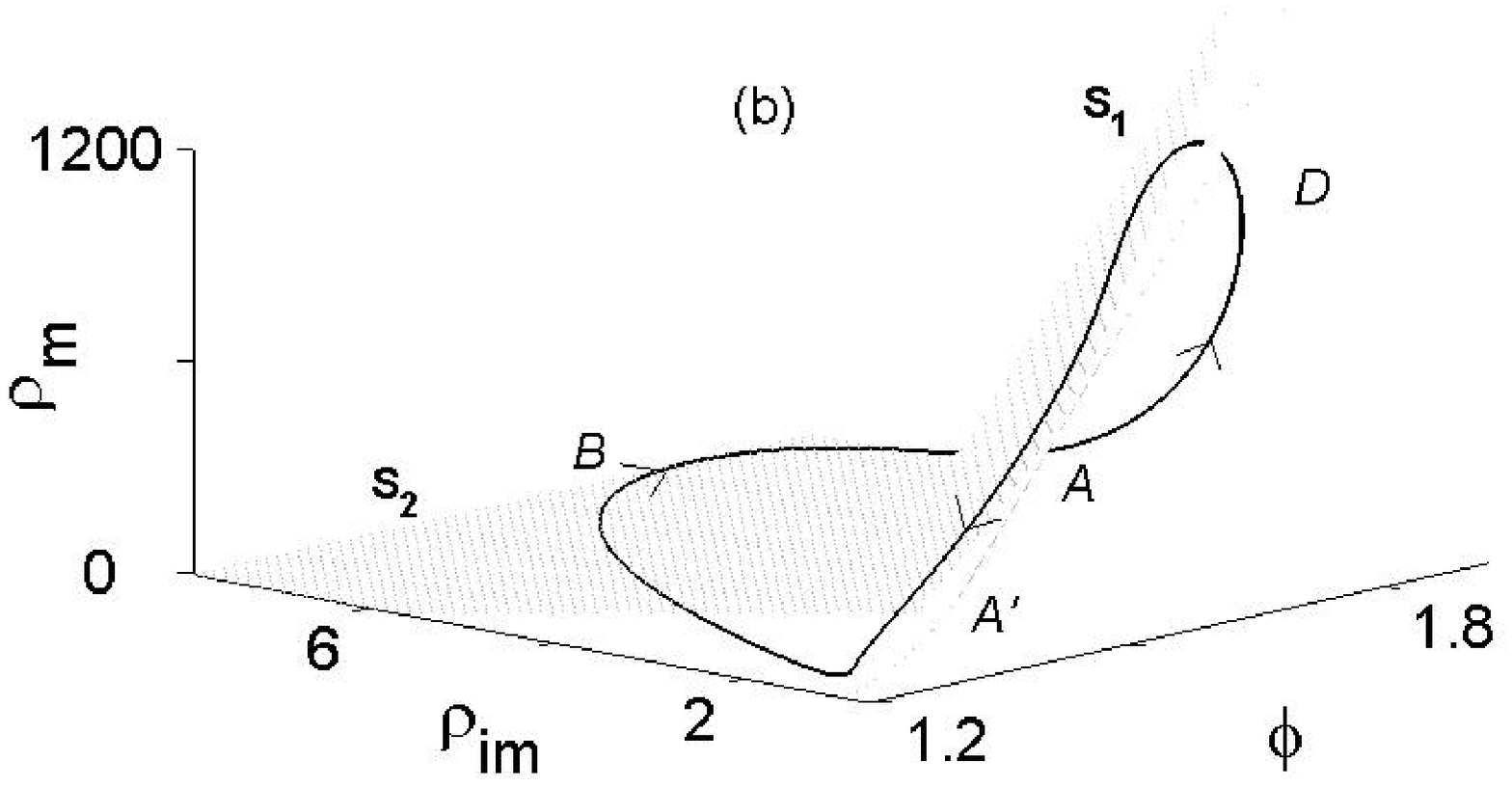}
\caption{(a)Bent slow manifold $S_1$ and $S_2$ (thick lines) with
a simple trajectory for $\dot\epsilon=200$ and $m= 3$. Inset:
$\rho_m$ (dotted curve) and $\phi$ (solid line).(b)The same trajectory in the
$(\phi,\rho_{im},\rho_m)$ space.} \label{fig12}
\end{figure}

Recently, the geometry of the slow manifold of the original model
has been analyzed in detail \cite{Rajesh00,Rajesh}. The analysis
shows that the relaxational nature of the PLC effect arises from
the atypical bent nature of the manifold. Here we recall some
relevant results on the slow manifold of the original model
($D=0$) and extend the ideas to the situation when the spatial
degrees of freedom are switched on ($D \neq 0$). Slow manifold
expresses the fast variable in terms of the slow variables,
conventionally done by setting the derivative of the fast variable
to zero \cite{Rajesh,Rajesh00}
\begin{equation}
\dot \rho_m = g(\rho_m,\phi) = - b_0 \rho_m^2 + \rho_m \delta
+\rho_{im} =0.
\label{slow1}
\end{equation}
where $\delta = \phi^m - \rho_{im} -a$. The variable $\delta$ has
been shown to have all the features of an effective stress and
thus plays an important physical role \cite{Rajesh00},
particularly in studying the pinning-unpinning of dislocations. We
note that $\delta$ is a combination of two slow variables $\phi$
and $\rho_{im}$ both of which  take small positive values. Hence,
$\delta$ takes on small positive and negative values. Using Eq.
\ref{slow1}, we get two solutions
\begin{equation}
\rho_m = [\delta + (\delta^2 + 4 b_0 \rho_{im})^{1/2}]/2b_0,
\end{equation}
one for $\delta <0 $ and another $\delta > 0$. For regions of
$\delta < 0$, as $b_0$ is small $\sim 10^{-4}$, we get
$\rho_m/\rho_{im} \approx - 1/\delta$ which takes on small values.
This defines   a part of the slow manifold, $S_2$ where $\rho_m$
is small. In this region, as the mobile density is small and
immobile density is large ( relative to $\rho_m$), this region can
be identified with pinned configuration of dislocations and hence
we shall refer to the region $S_2$ as the {\it 'pinned state of
dislocations'}. We note that  larger negative values of $\delta$
correspond to strongly pinned configurations, as they refer to
smaller ratio of $\rho_m/\rho_{im}$. For positive values of
$\delta$, another connected piece $S_1$ is defined by {\it large
values} of $\rho_m$, given by $\rho_m \approx \delta/b_0$, which
we refer to as the {\it 'unpinned state of dislocations'} as
$\rho_{im}$ is also small. These two pieces $S_2$ and $S_1$ are
separated by $\delta = 0$, which we refer to as {\it the fold
line} \cite{Rajesh,Rajesh00}(see below). A plot of the slow
manifold in the $\delta-\rho_m$ plane is shown in Fig.
~\ref{fig12}a. For the sake of illustration, we have plotted a
monoperiodic trajectory describing the changes in the densities
during a loading-unloading cycle.  The inset shows $\rho_m(t)$ and
$\phi(t)$. For completeness, the corresponding plot of the slow
manifold in the $(\rho_m,\rho_{im},\phi)$ space is shown in Fig.
~\ref{fig12}b, along with the trajectory and the symbols. In this
space, one can see that  $\delta =\phi^m-\rho_{im} -a = 0$ is a
line that separates the pieces  $S_2$ and $S_1$ of the slow
manifold, and hence the name {\it fold line}. The cyclic changes
in the variables is well captured by the nature of trajectory
shown in Fig. ~\ref{fig12}b. The trajectory enters $S_2$ at $A$
and moves into $S_2$, the value of $\delta$ ( in Fig.
\ref{fig12}a) decreases from zero to a maximum negative value as
the trajectory reaches $B$. Then $\delta$ increases as the
trajectory returns to $A^{\prime}$ before leaving $S_2$. The
corresponding segment is $ABA^{\prime}$ in Fig. ~\ref{fig12}b,
which is identified with the flat region of $\rho_m(t)$ in the
inset of Fig. ~\ref{fig12}a. As the trajectory crosses $\delta
=0$, $\partial g/\partial \rho_m $ becomes positive and it
accelerates into the shaded region (Fig. ~\ref{fig12}a) rapidly
till it reaches $\rho_m = \delta/2b_0$. Thereafter it settles down
quickly on $S_1$ decreasing rapidly till it reenters $S_2$ again
at $A$. The burst in $\rho_m$ (inset in Fig. ~\ref{fig12}a)
corresponds to the segment $A^{\prime}DA$ in Fig. ~\ref{fig12}a
and b. The nature of trajectories for higher strain rate remains
essentially the same, but is chaotic. The nature of the
trajectories in the power law regime of strain rates which will be
discussed later.

Having identified the regions of the slow manifold with  the
pinned and unpinned states of dislocations, we now  consider the
variation of stress when dislocations are pinned and are unpinned.
Consider the stress changes as the state of the system goes though
a burst of plastic activity. For $D =0$, Eq. \ref{Eq: seqn}
reduces to
\begin{equation}
\dot \phi = d[ \dot \epsilon -\dot {\epsilon}_p],
\end{equation}
where $\dot\epsilon_p = \phi^m \rho_m$ defines the plastic strain
rate. Since $\rho_m$ is small and nearly constant on $S_2$, stress
increases monotonically. However, during the burst in $\rho_m$
($A^{\prime} D A$ in the inset), $\dot \epsilon_p (t)$ exceeds
$\dot \epsilon$ leading to an yield drop. Since $\rho_m$ grows
outside $S_2$, $\delta =0$ line separates the pinned state from
the unpinned state. Thus, $\delta =0$ {\it physically corresponds
to the value of the effective stress at which dislocations are
unpinned.}

When the spatial degrees of freedom are included, there is no
additional complication as the slow manifold is defined at each
point. In this case, a convenient set of variables for
visualization of dislocations is $(\rho_m (x), \delta (x), x)$.
Here, our aim is to investigate the nature of  typical spatial
configurations in the chaotic and the power law regimes of stress
drops and study the changes as we increase the strain rate. For
simplicity, we shall use $h=0$ for which we have
$\phi_{eff}=\phi$. ( It is straightforward to extend the arguments
to the case when $h \ne 0$.) Then, the plastic strain rate $\dot
\epsilon_p(t)$ is given by
\begin{equation}
\dot \epsilon_p(t) = \phi^m(t)\frac{1}{l}\int_0^l\rho_m(x,t)dx =
\phi^m(t) \bar\rho_m(t)\label{marstab},
\end{equation}
where $\bar \rho_m(t)$ is the mean mobile density
($=\sum_j\rho_m(j,t)/N$ in the discretized form).  With the
inclusion of spatial degrees of freedom, the yield drop is
controlled by the spatial average $\bar \rho_m(t)$ rather than by
individual values of $\rho_m(j)$. Further, we note that the
configuration of dislocations change during one loading-unloading
cycle. However, one should expect that configurations will be
representative for a given strain rate. Further, we know that the
drastic changes occurs during an yield drop when $\bar \rho_m(t)$
grows rapidly. Thus, we focus our attention on the spatial
configurations on the slow manifold at the onset and at the end of
typical yield drops.

\begin{figure}
\includegraphics[height=8cm,width=8.8cm]{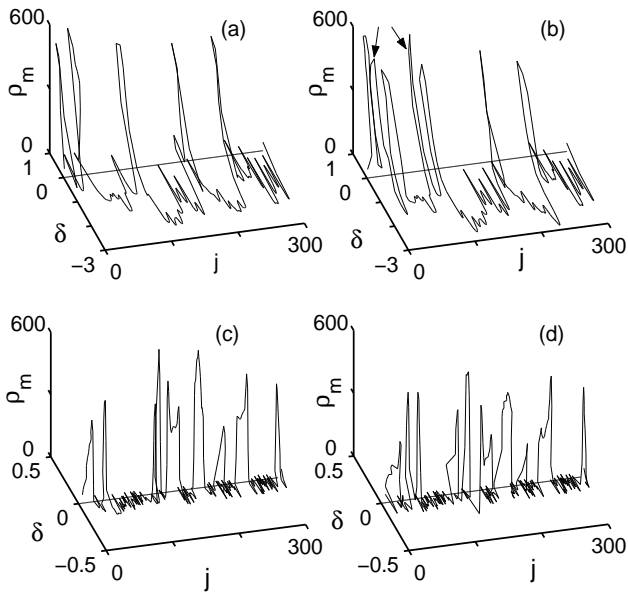}
\caption{Dislocation configurations on the slow manifold at the
inset and at the end of an yield drop: (a) and (b) for
$\dot\epsilon=120$ (chaotic regime), and (c) and ( d) for
$\dot\epsilon=280$ (scaling regime).} \label{fig13}
\end{figure}

First consider the configuration seen just before and after the
yield drop when the strain rate is in the chaotic regime. In this
regime, the stress drop magnitudes are large which implies that
the change in mobile density is large. Figures \ref{fig13} a, b
for a typical value of $\dot \epsilon = 120$. It is clear that
both at the onset and at the end of a typical large yield drop,
the $\delta(j)$ values which reflect the state of system ( pinned
or unpinned state), is negative and correspondingly  the mobile
density $\rho_m(j)$'s are small, i.e., most dislocations are in
{\it a strongly pinned state}. ( Recall that $\delta$  signifies
how close the spatial elements are close to unpinning threshold.)
The arrows show the increase in $\rho_m(j)$ at the end of the
yield drop. We have checked that this is a general feature for all
yield drops in the chaotic regime of strain rates. Now consider
dislocation configuration in the scaling regime at high strain
rates, say, $\dot \epsilon =280$, at the onset and at the end of
an yield drop shown in Fig. ~\ref{fig13} c,d respectively. In
contrast to the chaotic regime, in the scaling regime, {\it most
dislocations are clearly seen to be at the threshold of unpinning
with $\delta(j) \approx 0$,} both at the onset and end of the
yield drop. This also implies that they remain close to this
threshold all through the process of an stress drop. We have
verified that {\it the edge-of-unpinning picture is valid} in the
entire power law regime of stress drops for a range of $N$ values.
Further, as a function strain rate, we find that  the number of
spatial elements reaching the threshold of unpinning $\delta = 0$
during an yield drop increases as we approach the scaling regime.

\section{TYPES OF BANDS}

The fact that the spatially extended Ananthakrishna's  model is
able to successfully reproduce the crossover dynamics from chaos
to the  power-law regime of stress drops ( and other generic
features demonstrated earlier) might suggest that the
characteristic features of the PLC bands may also emerge out of
the model. Most models of dislocations bands use diffusive
coupling although the physical mechanism of the term is different
in different situations \cite{KFA}. An important feature of the
spatial coupling in the model is that it accounts for spreading of
dislocations into regions of low back stress once dislocations are
unpinned ( the factor $\rho^{-1}_{im}$). The term also determines
the length scale over which dislocations spread into the
neighboring elements. Thus, while dislocation pinning and
unpinning gives a heterogeneity in space (in principle), regions
of low $\rho_{im}$ are favored for dislocation multiplication and
spreading into neighboring regions. Further, this type of spatial
term couples length scale and time scales in a dynamical way as
$\rho_{im}$ itself evolves in time and hence the associated time
scale. Indeed, multiplication of dislocation depends on stress,
({\it i.e.,} $\phi_{eff}^m$), and hence this rate itself is
changing dynamically leading to changes in the time scale of
internal relaxation as a function of $\dot \epsilon$. We expect
this to lead to changes in spatial correlation as strain rate is
increased.

Below we report both numerical and analytical studies on the
spatiotemporal patterns emerging from the model as a function of
the strain rate, $\dot\epsilon$. We begin with the numerical
results \cite{Bhar03a}.

For $\dot\epsilon < 10$ and $\dot\epsilon > 2000$, we get
homogeneous steady state solutions for all the dislocation
densities, $\rho_m$, $\rho_{im}$ and $\rho_c$. In these ranges of
strain rates, $\phi$ takes the fixed point values asymptotically.
In the region where interesting dynamics of chaotic and power law
states are observed, the nature of the dislocation bands can be
broadly classified into three different types occurring at low,
intermediate and high strain rates described below.

\begin{figure}
\includegraphics[height=6cm,width=8cm]{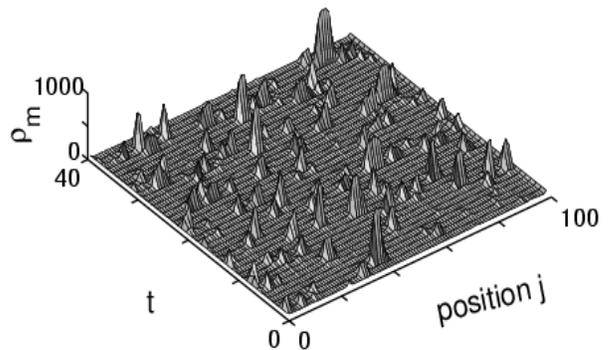}
\caption{Spatially uncorrelated bands at $\dot\epsilon=40$.}
\label{fig14}
\end{figure}

For strain rates, $30\le \dot\epsilon <70$, we get uncorrelated
static dislocation bands.  The features of these bands are
illustrated for a typical value, say for $\dot\epsilon=40$. A plot
of $\rho_m(j,t)$ is given in Fig.~\ref{fig14}. Dislocation bands
of finite width nucleate randomly in space and they remain static
till another band is nucleated at another spatially uncorrelated
site. The associated stress-time curves  which are nearly regular
have large characteristic stress drops. The distribution of these
stress drops is found to be peaked as in experiments at low strain
rates \cite{Bhar01}.

\begin{figure}
\includegraphics[height=6cm,width=8cm]{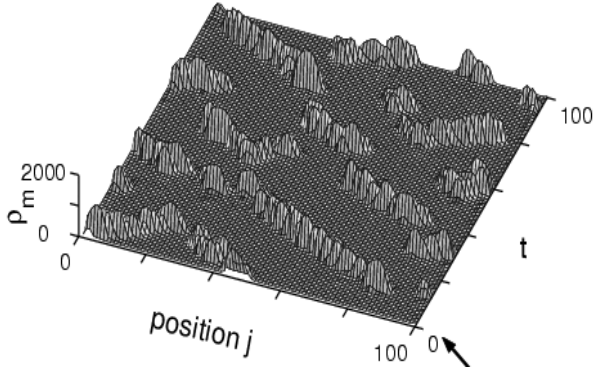}
\caption{Hopping type bands at $\dot\epsilon=130$ (arrow shows one
such band).} \label{fig15}
\end{figure}

At slightly higher values of strain rates, $70 \le \dot\epsilon <
180$ we find that new bands nucleate ahead of the earlier ones,
giving a visual impression of {\it hopping bands}. This can be
clearly seen from Fig.~\ref{fig15} where a plot of $\rho_m(j,t)$
is given for $\dot\epsilon=130$. However, this hopping motion does
not continue till the other boundary. They stop midway and another
set of hopping bands reappear in the neighborhood. Often
nucleation occurs at more than one location. Stress-time plots in
this regime have a form similar to Fig. ~\ref{fig1}c with the
average amplitude of the stress drops being smaller than the
localized non-hopping bands at low strain rates as seen in
experiments. These stress drops also have a nearly symmetric
peaked distribution as in the previous case but slightly skewed to
the right similar to those observed in experiments \cite{Bhar01}.

\begin{figure}
\includegraphics[height=6cm,width=8cm]{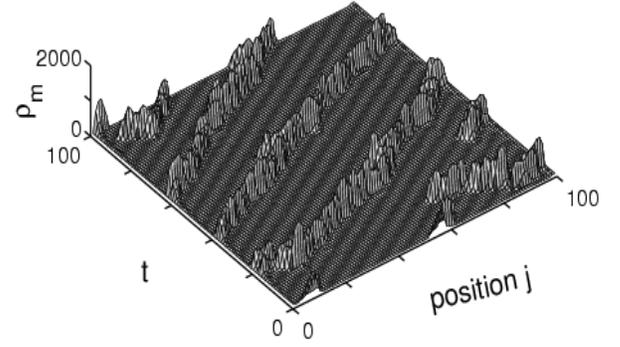}
\caption{Fully propagating  bands at $\dot\epsilon=240$}
\label{fig16}
\end{figure}

As the strain rate is increased further, the extent of propagation
increases, concomitantly, the magnitudes of the stress drops
decrease. We  see continuously propagating bands even at $\dot
\epsilon =240$ as can be seen from Fig. \ref{fig16}. One can see
dislocation bands nucleating from one end of the sample ($j=0, \,
t=25,50$ and 75) and propagating continuously to the other end.
Often, we see a band nucleating at a point, branching out and
propagating only partially towards both the ends. Unlike the
present case which exhibits rather uniform values of $\rho_m$, we
usually find irregularities as the band reaches the edges. The
stress strain curves in this region of strain rates, exhibit scale
free feature in the amplitude of the stress drops (Fig. \ref{fig1}
d) with a large number of small drops. As can be seen from  Fig.
\ref{fig1} d, the mean stress level of these small amplitude
stress drops increases until a large yield drop is seen. This
large stress drop corresponds to bands having reached the end of
the specimen.

\begin{figure}
\includegraphics[height=5cm,width=8cm]{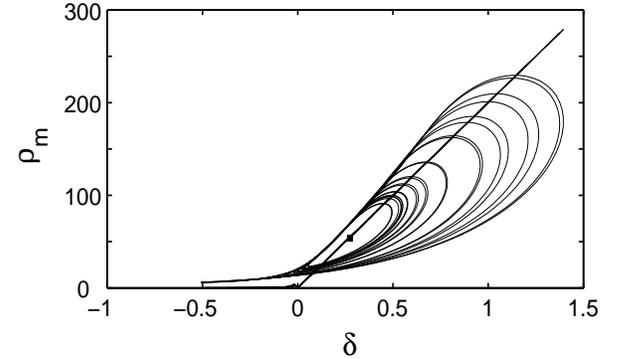}
\caption{Slow manifold showing a trajectory for the space
independent model near the reverse Hopf bifurcation point, at
$\dot\epsilon=90$, $m=2$. $\bullet$ fixed point of Eqn 1-3 and 6.} \label{fig17}
\end{figure}

It is possible to calculate the velocity of the propagating bands
in the high strain rate limit. We first note that our equations
constitute a coupled set of integro-partial differential
equations, and hence cannot be dealt with in their present form.
To reduce these equations to a  form that is suitable for further
analysis, we recall a few pertinent points about the changes in
the structure of the slow manifold as a function of the applied
strain rate. We note that the original model exhibits an
incomplete approach to homoclinicity \cite{Rajesh00}, {\it i.e.,}
the number of mixed mode oscillations of the type $L^s$, where $s$
refers to the small period oscillations and $L$ refers to a large
relaxation oscillation, are limited. Typically, about 12 small
period nearly harmonic oscillations are known to occur for a
single large one at high values of the strain rate.  The reason
attributed to this is the finite rate of softening of the eigen
value of the fixed point due to presence of the reverse Hopf
bifurcation is reached \cite{Rajesh00}. In the presence of the
spatial coupling we find that the softening is further enhanced as
is clear from the fact that the upper Hopf bifurcation is pushed
to much larger values of strain rate ( $\dot\epsilon_{c_2}=2000$,
see Fig. \ref{fig0}). This enhanced softening rate implies that
the number of small period oscillations is also increased in this
domain of strain rates. Even so, the geometry of the slow manifold
is not altered from that of the space independent model. In
particular, the position of the unstable saddle focus remains
located on the $S_1$ part of the manifold
(Ref.\cite{Rajesh00,Rajesh}). In addition, the feature of the
fixed point approaching the fold line as a function of the strain
rate is retained. Under these conditions, for high strain rates
nearly sinusoidal oscillations are executed around the fixed point
with the orbits touching $S_2$ only after executing several such
turns. A plot of this is shown in Fig. ~\ref{fig17}.
\noindent To understand the dynamics at high strain rates, we
recall that our analysis in Ref. \cite{Rajesh00}  shows that the
orbit is re-injected  along the stable manifold  close to the
unstable saddle focus ( as shown in Fig.11 of
Ref.\cite{Rajesh00}). The orbit then spirals out along the
unstable manifold of the fixed point. Once the orbit is
sufficiently away from the fixed point when the influence of the
fixed point is lost,  it is re-injected close to the fixed point
via $S_1$. The dynamics then repeats. Note that at high applied
strain rate, the system is close to reverse Hopf bifurcation point
and hence the fixed point is close to the applied strain rate
value. Thus as the orbit executes one turn, there is one small
yield drop. However, the orbit executes several turns around the
fixed point, each turn leading larger loop sizes, i.e., larger
values of $\rho_m$ and consequently to successively larger stress
levels than the earlier one  before briefly visiting $S_2$.

Under these conditions the dynamics is entirely controlled by the
spiralling motion around the fixed point. Thus, the entire
dynamics is essentially described by the fast variable; the other
two variables $\rho_{im}$ and $\phi$ can be taken to be
parameters. Such a situation is described by the transient
dynamics dictated entirely by equation of the fast variable (the
so called layer problem \cite{Milik}) and thus, we are justified
in using only the evolution equation of the fast variable in terms
of the slow manifold parameter $\delta = \phi^m-\rho_{im}-a$.
Since the trajectory rarely visits the $S_2$ part of the slow
manifold, we restrict the calculations to $\delta
> 0$. The physical picture of a propagating solution is that as
the orbit at a site makes one turn around the fixed point, {\it
i.e.,} $\delta $ small but positive, around the value of the
applied strain rate, the front advances by a certain distance
along the specimen like the motion of a screw.

The rate equation for the mobile dislocation density
$\rho_m$ in terms of $\delta$ is
\begin{equation}
\frac{\partial\rho_m}{\partial t} = -b_o\rho_m^2 + \delta \rho_m +
\rho_{im} + D'\frac{{\partial}^2\rho_m}{\partial x^2},
\label{marglin}
\end{equation}
where $D'=D\phi^m/\rho_{im}$. Since, the slow variables,
$\rho_{im}$ and $\phi$ are treated as parameters, this has the
form of Fisher-Kolmogorov equation for propagating fronts, which
has been well studied. This equation can be reduced to the
standard form
\begin{equation}
\frac{\partial Z}{\partial t'} =  Z (1-Z)  + D'\frac{{\partial}^2
Z}{\partial x^2}, \label{marglin1}
\end{equation}
(This is done by first transforming $\rho_m = X -
\rho_{im}/\delta$, dropping the term $2b_0\rho_{im}/\delta$
compared to $\delta$ in the linear term in $X$, and then using $ Z
= X \delta/b_0 $ and $t' = t \delta$.) It is clear that $Z=0$ is
unstable and $Z =1$ is stable. Using the form for propagating
front $Z=Z_oe^{\omega t' -kx'}$, the marginal velocity is
calculated using $v^* = Re \omega (k^*)/Re k^* = d\omega/d
k\vert_{k = k^*}$ and $\Im d\omega/d k\vert_{k = k^*} = 0$, gives
the velocity of the bands $v^*=2$ \cite{Dee,Saarloos}. In terms of
the variables in Eqn.~\ref{marglin}, the marginal velocity is
\begin{equation}
v^*=2\sqrt{D\delta}.
\label{marglvel}
\end{equation}
\noindent In order to relate this to the applied strain rate, we
note that  for a fixed value of the strain rate (where propagating
bands are seen), the average level of stress drop is essentially
constant. Thus, from Eqn.~\ref{Eq: seqn}, we see that in this
regime of high strain rates, the applied strain rate
$\dot\epsilon$ is essentially balanced by the plastic strain rate
$ ({1}/{l})\int_0^l\phi^m\rho_m(x,t) \equiv \dot{\epsilon}_p$.
Then, using $\phi^m = \dot \epsilon / \bar {\rho}_m$,  and using
$\delta = \phi^m - a  -\rho_{im}$, we get
\begin{equation}
v=2\sqrt{\frac{D\dot\epsilon}{\bar{\rho}_m\rho_{im}}\big(\frac{\dot\epsilon}{\bar{\rho}_m}-a-\rho_{im}\big)}.
\label{velarg2}
\end{equation}
It is important to note that at high applied strain rate
$\bar\rho_m \sim \bar\rho_m^*$ , the fixed point value. Thus, for
all practical purposes, we can assume $\bar\rho_m$ as a constant.
From the above equation, we see that the velocity of the
propagating bands is proportional to the applied strain rate. This
result is similar to the result obtained recently by H\"ahner {\it
et al.} \cite{Hahner02}. Further, $v \propto \bar \rho_m^{-1}$
which also appears to be consistent with an old experimental
result. (See Fig. 7 of Ref. \cite{Korbel} which appears to fit $v
\bar \rho_m$ = constant.) This result needs further experimental
support.

As the form of our equation has been reduced to the standard form,
all other results carry through, including nonlinear analysis. We
have numerically calculated the velocity of the continuously
propagating bands at high strain rates from the model which
confirms the linear dependence of the band velocity on applied
strain rate. In the region of strain rates $\dot\epsilon =$ 220 to
280 (corresponding to unscaled strain rate values $10^{-5} - 1.5
\times 10^{-5}s^{-1}$), we find that the unscaled values of the
band velocity increases from 100 to 130 $\mu m/s$. These values
are consistent with the experimental values reported by H\"ahner
et. al. \cite{Hahner02}.

We note here that the types of the bands seen in our model are
correlated with the two distinct dynamical  regimes investigated.
The hopping type bands belong to the chaotic regime, a result
consistent with the recent studies on Cu-Al polycrystals
\cite{Bhar01}.  On the other hand, the propagating bands are seen
in the power law regime of stress drops \cite{Bhar02}, again
consistent with these studies \cite{Anan99,Bhar01}. Curiously the
uncorrelated bands predicted by the model also belong to the
chaotic regime. We shall now explain these results based on the
dynamics of the model. We first note that each spatial element is
described the three dislocation densities ( Eq. (1-3)). Consider
one of these elements being close to unpinning threshold, ie.,
$\delta =0$. It has been shown earlier that $\rho_{im}$ is out of
phase with $\rho_m$ \cite{Rajesh,Rajesh00}. This feature is
retained with the spatial coupling as well.  When the orbit is
about to leave $S_2$, ie., when $\rho_m(j)$ is at the verge of a
sharp increase, $\rho_{im}$ is largest. However, the extent of the
spatial coupling is determined by $\rho_{im}^{-1}$. But the
magnitude of $\rho_{im}$ itself decreases with the applied strain
rate, being large at low strain rates \cite{Rajesh,Rajesh00}.
Thus, the spatial width of this is small at low $\dot \epsilon$
and large at high $\dot \epsilon$. Next we note that the growth
and decay of $\rho_m(j)$ with $j$ occurs over a short time scale
which is typically of the order of the correlation time, $\tau_c$,
of $\phi(t)$. Beyond this time, the memory of its initial state is
lost. Consider an initial state when a band is formed at some
location. Before the memory of this initial state decays, if a new
band is {\it not} created, we get an uncorrelated band. On the
other hand, if a new band is created before the memory of the
initial state decays, there are two possibilities. If another band
is created just before the correlation decays substantially  by
that time, we get a hopping type band. If however, even before the
burst of $\rho_m (j)$ decreases beyond its peak value, new sources
of creation of $\rho_m$ occur, then we end up seeing a propagating
band. An analysis of  the correlation time shows that it increases
with the applied strain rate. Concomitantly, $\rho_{im}$ decreases
with $\dot\epsilon$ which implies that the spatial correlation
increases. (Indeed, the value $\rho_{im}$ is quite small for large
$\dot\epsilon$ as we reach the power law regime of stress drops.)
Under these conditions, only partial plastic relaxation is
possible in this regime. This discussion clarifies the dynamic
interplay of time scales and length scales. Moreover, as the
spatial coupling term allows the spreading of dislocations only
into regions of low $\rho_{im}$ or low back stress, the propensity
for continuous propagation of the band is enhanced when
$\rho_{im}$ is small. In addition, we find that higher values of
$\rho_{im}$ at the wake of the band which favors propagation into
regions of smaller immobile density thus determining the direction
of propagation also.

\section{Summary and Conclusions}

We first summarize and make appropriate comments wherever
necessary.  Detailed numerical and analytical studies on the
extended Ananthakrishna's model shows that it reproduces all the
important features of the PLC effect including  the crossover in
the dynamics from a chaotic to a power law regime observed in
experiments. It also provides insight into the dynamical causes
leading to this crossover. A systematic study of the system size
effects of the Lyapunov spectrum carried out elucidates the
underlying mechanism controlling the crossover. The study
demonstrates that the limiting Lyapunov distribution evolves from
a set of positive and negative exponents with a few null exponents
to a dense set of null exponents as we approach the scaling regime
of stress drops. This study is complemented through an analysis of
the slow manifold. This method  is particularly useful in giving a
geometrical picture of the spatial configurations, both in the
chaotic and power law regime of stress drops. The study shows that
{\it the configuration of dislocations is largely in the pinned
state in low and medium strain rates (chaotic domain) are pushed
to  the threshold of unpinning as we increase strain rate ( power
law stress drop regime)}.  The study also establishes that the
present model has considerable similarities with the GOY model of
turbulence \cite{Yam87}.  The model also reproduces the major
spatial features of the PLC effect. The randomly nucleated band,
the hopping and propagating types are found as the strain rate is
increased. It also predicts a linear dependence of the velocity of
the band and inverse dependence on the mobile density at high
strain rates.

Several observations  may be in order on the dynamics of the
crossover. We first note that the crossover itself is smooth as
the changes in the Lyapunov spectrum are gradual, though it occurs
in a narrow interval of  strain rates ( from 220-250). Second, the
power law here is of purely dynamical origin (in the sense
elaborated below below). We have shown that this is a direct
result of the existence of a reverse Hopf bifurcation at high
strain rates. In this regime due to softening of the eigen values
( as a function of the applied strain rate), the orbits are mostly
restricted to the region around the saddle node fixed point
located on the $S_1$ part of the manifold. This offers a dynamical
reason for the smallness of the yield drops in this region
\cite{Rajesh,Rajesh00}.    Note also that there is a dynamic feed
back between the stress determined by Eq. \ref{Eq: seqn} and the
production of dislocations in Eq. \ref{Eq: xeqn} which provides an
explanation for the slowing down of the plastic relaxation. ( The
partial plastic relaxation has been cited as the reason for the
power law \cite{Anan99,Bhar01}.) This sets up a competition
between the time scale of internal relaxation and the time scale
determined by the applied strain rate ( essentially Deborah
number). We note that while the time scale for internal relaxation
is increasing, that due to the applied strain rate is decreasing.
Third, our analysis shows that the power law regime of stress
drops occurring at high strain rates belongs to a different
universality class compared to SOC systems, as it is characterized
by a dense set of null exponents.  This must be contrasted with
the lack of any characteristic feature of the nature of the
Lyapunov spectrum in the few models of SOC studied so far
\cite{Cess,desouza,Erzan}. For instance, no zero and positive
exponents, zero exponent in the large $N$ limit etc., have been
reported \cite{Cess,desouza,Erzan}. ( Often, the nature of largest
Lyapunov exponent is inferred based on the similarity of other
dynamical invariants \cite{Erzan}.) The dense set of null
exponents in our model is actually similar to that obtained in
shell models of turbulence where the power law is seen at high
drive values \cite{Yam87}. However, there are significant
differences. First, we note that the shell model \cite{Yam87}
cannot explain the crossover as it is only designed to explain the
power law regime. Second, the maximum Lyapunov exponent is large
for small viscosity parameter $\eta$, ie., $\lambda_1 \propto
\eta^{-1/2}$ in shell models \cite{Yam87} in contrast to near zero
value in our model. It is also interesting to note that in our
model  propagating solutions arise in the power law regime of
stress which comes as a surprise. As far as we are aware, this is
the first situation, both from a experimental and theoretical
angle, where propagating solutions are seen  in  a marginally
stable situation.

Regarding the spatial features seen in the model, we stress that
these features emerge purely due to dynamical reasons without any
recourse to using  the negative strain rate sensitivity feature as
an input, as is the case in most models
\cite{Hahner02,Mc,Zhang,Leby}. Even the recently introduced
poly-crystalline plasticity model which reproduces the crossover
behavior also uses the negative SRS as an input \cite{Kok}. The
dynamical approach followed here clearly exposes how the slowing
down of the plastic relaxation occurs {\it due to a feed back
mechanism of dislocation multiplication and applied strain rate}
as we reach the power law regime of stress drops. While the three
different types of bands have features of the uncorrelated type C,
hopping type B and the propagating type A bands  found in
poly=crystalline materials, there is no element of
poly-crystallinity in the model in its present form. In
poly-crystals, other types of coupling terms do arise which have
also been modelled by  diffusive type terms \cite{KFA}. One way of
including the presence of grain boundaries within the natural
setting of the model is to recognize that crossslip will be
hindered near the grain boundaries which also leads to a term
similar to the present diffusive term. Such a term can account for
the back stress arising from the incompatibility of grains. As the
form of these terms  are similar, the basic results are unlikely
to change although one should expect a competition between the
terms operating within a crystal and that at the grain boundaries.

From a purely dynamical point of view, this model should be of
interest to the area of dynamical systems, as it appears to be the
first fully dynamical model which exhibits a crossover from a
chaotic to a power law regime,  in the sense that our model is
continuous space time model {\it without any recourse to
artificial thresholds} as is done in coupled map lattices
\cite{Chate}. We note also that while the slow manifold subspace
gives a method of visualizing the dislocations configurations,
particularly in the scaling regime, the complementary subspace of
the fast variable has helped us to obtain the band velocity in the
same regime of strain rates. From the point of view of plastic
instabilities, the present dynamical approach should be a
promising direction for explaining many other patterns mentioned
in the introduction \cite{KFA}.

Finally, as stated in the introduction, the PLC effect bears
considerable similarity with many stick-slip systems and hence
presents a way of understanding some of these systems. Here we
make a few comments on the similarity of the present model for a
possible adoption to the observed voltage fluctuations in charge
density wave (CDW) compounds \cite{Dumas,Anan98}. Under the action
of applied electric field, anomalously large voltage fluctuations
are reported when the electric field is above the threshold value.
This ohmic to non-ohmic transition in $K_{0.3}MoO_3$ and
$Rb_{0.3}MoO_3$, for instance \cite{Dumas1}, has not been
adequately explained although the similarity with the PLC effect
has been noted \cite{Dumas}. Lee and Rice \cite{Lee} have
suggested that phase dislocations of the CDW carry current at
fields too low for the CDW  to move as a whole. Indeed both fall
in the category of pinning-depinning phenomenon. In the case of
CDW, pinned at impurity/defect sites is unpinned due to the
applied electric field. The threshold value of the electric field
can be viewed as the onset of plastic flow of the charged phase
dislocations \cite{Dumas}. These authors  identify stress with
voltage and strain with current and suggest that the total current
is the sum of the ohmic part and that arising from charge density
waves. This is corresponds to the elastic and plastic
displacements competing to give rise to the PLC instability. We
believe that this parallel can be taken further along the lines of
our PLC model where one can identify the phase dislocations with
mobile dislocations, the neutral defects of the CDW with
dislocation dipoles, ie., the immobile, and the phase dislocations
pinned at defects with the Cottrell type. Work along these lines
is progress.

This work is supported by Department of Science and Technology,
New Delhi, India.

\end{document}